\documentclass[onecolumn]{emulateapj}
\usepackage{amssymb,amsfonts,amsmath,amstext,amsgen,amsopn,amsxtra,indentfirst,lscape}

\newcommand{\gtsim}{{\gtrsim}}
\newcommand{\ltsim}{{\lesssim}}
\newcommand{\poro}{{\cal P}}
\newcommand{\poromax}{0.55}
\newcommand{\RX}{{\cal R}_{\rm X}}
\newcommand{\RXobs}{{\cal R}_{\rm X,obs}}
\newcommand{\TX}{{\cal T}_{\rm X}}
\newcommand{\TXobs}{{\cal T}_{\rm X,obs}}
\newcommand{\sigmax}{{\Delta\sigma}}
\shorttitle{Fluffy Grain Aggregates}
\shortauthors{Heng \& Draine}
\begin{document}

\title{Constraining the Porosities of Interstellar Dust Grains}

\author{Kevin Heng\altaffilmark{1} \& Bruce T. Draine\altaffilmark{2}}

\altaffiltext{1}{Institute for Advanced Study, School of Natural
  Sciences, Einstein Drive, Princeton, NJ 08540; heng@ias.edu}
\altaffiltext{2}{Princeton University Observatory, Peyton Hall,
  Princeton, NJ 08544; draine@astro.princeton.edu}

\begin{abstract}
We present theoretical calculations of the X-ray scattering properties
of porous grain aggregates with olivine monomers.  The small and large
angle scattering properties of these aggregates are governed by the
global structure and substructure of the grain, respectively.  We
construct two diagnostics, $\RX$ and $\TX$, based on the optical and
X-ray properties of the aggregates, and apply them to a {\it Chandra}
measurement of the dust halo around the Galactic binary GX13+1.  Grain
aggregates with porosities ${\cal P} \gtrsim \poromax$ are ruled out.
Future high-precision observations of X-ray dust haloes together with
detailed modeling of the X-ray scattering properties of porous grain
mixtures will further constrain the presence of porous grain
aggregates in a given dust population.
\end{abstract}


\section{Introduction}
\label{sect:intro}

%figure 1: schematic
\begin{figure}
\begin{center}
\includegraphics[width=6.0in]{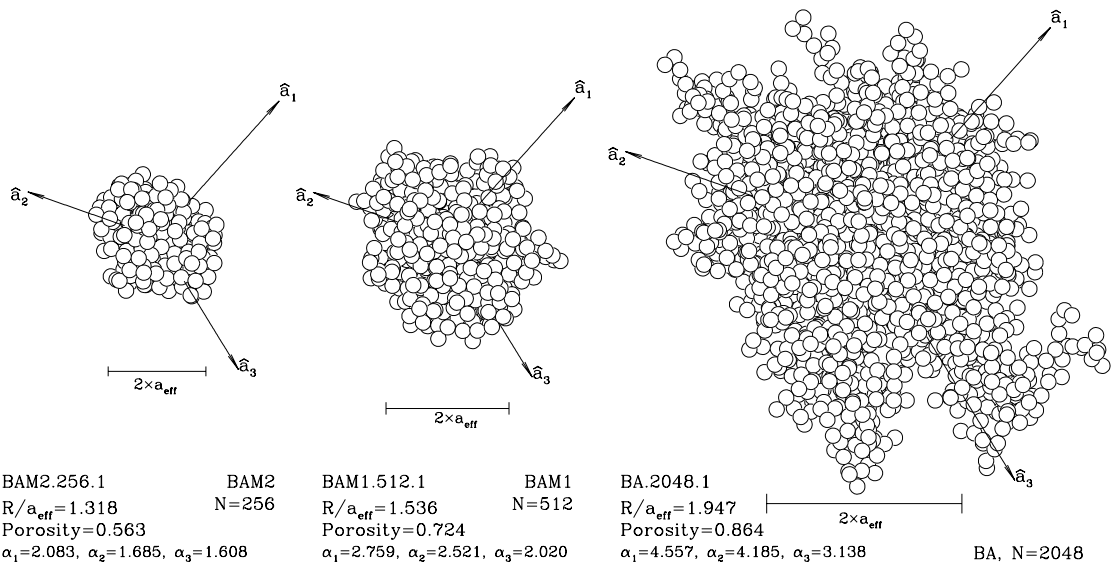}
\end{center}
\caption{
  Random aggregates produced by the BAM2, BAM1, and BA agglomeration
  schemes (Shen, Draine \& Johnson 2008).
  The unit vector $\hat{a}_1$ is
  along the principal axis of the largest moment of inertia.
  With $a_0=0.04\micron$ silicate spherules, each of these aggregates
  (with $a_{\rm eff}=0.254$, $0.320$, and $0.508\micron$) has
  $Q_{\rm ext}(B)/Q_{\rm ext}(R) \approx 1.70$, the observed value of
  $A_B/A_R$ (see text).
  }
\label{fig:schematic}
\end{figure}

After many decades of trying to deduce the properties of interstellar
dust from observations, the basic geometric structure of the grains
remains controversial.  The observed polarization of starlight
establishes that some or all grains are non-spherical, and there is
agreement that interstellar grains must include both silicate material
and carbonaceous material, but there are divergent views concerning the
internal structure of grains.  Some authors (e.g., Mathis et al.\ 1977;
Draine \& Lee 1984; Kim, Martin \& Hendry 1994; Kim \& Martin 1995;
Weingartner \& Draine 2001, hereafter WD01; Draine \& Li 2007; Draine
\& Fraisse 2009) have modeled interstellar grains as compact,
zero-porosity spheres or spheroids, with some consisting of silicate
and others consisting of carbonaceous material.  Others have argued
that the evolution of interstellar grains involves coagulation in
clouds, producing grains with a mixed composition and a ``fluffy''
structure, with porosities taken to be $\sim 0.8$ (Mathis \& Whiffen
1989) or $\gtrsim0.9$ (Voshchinnikov et al.\ 2006).  

Because polarized light-scattering is sensitive to the geometry of the
scatterers, it has been possible to use the observed color and polarization
of scattered light, as a function of scattering angle,
to estimate the size and porosity of cometary dust particles
and the dust in debris disks.
Shen et al.\ (2008, hereafter SDJ08) 
examined the optical properties of porous grain
aggregates using the discrete dipole approximation (Draine 1988;
Draine \& Flatau 1994).
Shen et al.\ (2009) found that moderate-porosity aggregates can
reproduce the observed scattering properties of dust in the AU Mic debris disk
and in comets.
Unfortunately, the scattering properties
of interstellar grains remain uncertain, and to date there has been no
observational diagnostic capable of discriminating between
these different models for interstellar grains.

The present study shows that X-ray scattering can discriminate
between compact and fluffy models for interstellar dust.
Compact grain models have been successful in reproducing observations
of X-ray scattering halos (e.g., Draine \& Tan 2003, Smith 2008).  
In the present paper, we show how the X-ray scattering properties of
irregular porous grains can be accurately calculated.  We find that
compact and fluffy grains differ substantially in their X-ray
scattering properties, because the X-rays can ``see'' the small-scale
structure within an aggregate grain, and because fluffy grains must be
larger than compact grains when required to reproduce observed
interstellar reddening constraints.

In \S\ref{sect:physics}, we describe the X-ray optics of grain
aggregates, including approximate scaling laws and our application of
anomalous diffraction theory (ADT) to calculate their X-ray scattering
properties.  We present the results of our calculations in
\S\ref{sect:results}.  In \S\ref{sect:diagnostic}, we construct two
diagnostics based on the optical and X-ray properties of the dust
grains and apply them to the {\it Chandra} halo measurement around the
Galactic binary GX13+1 by Smith (2008, hereafter S08).  We are able to
rule out grain aggregate models with porosities ${\cal P}\gtsim\poromax$.  
We summarize and discuss our results in \S\ref{sect:discussion}.

\section{X-ray Optics of Grain Aggregates}
\label{sect:physics}

%figure 2: 2D theoretical halo
\begin{figure}
\begin{center}
\includegraphics[width=3.2in]{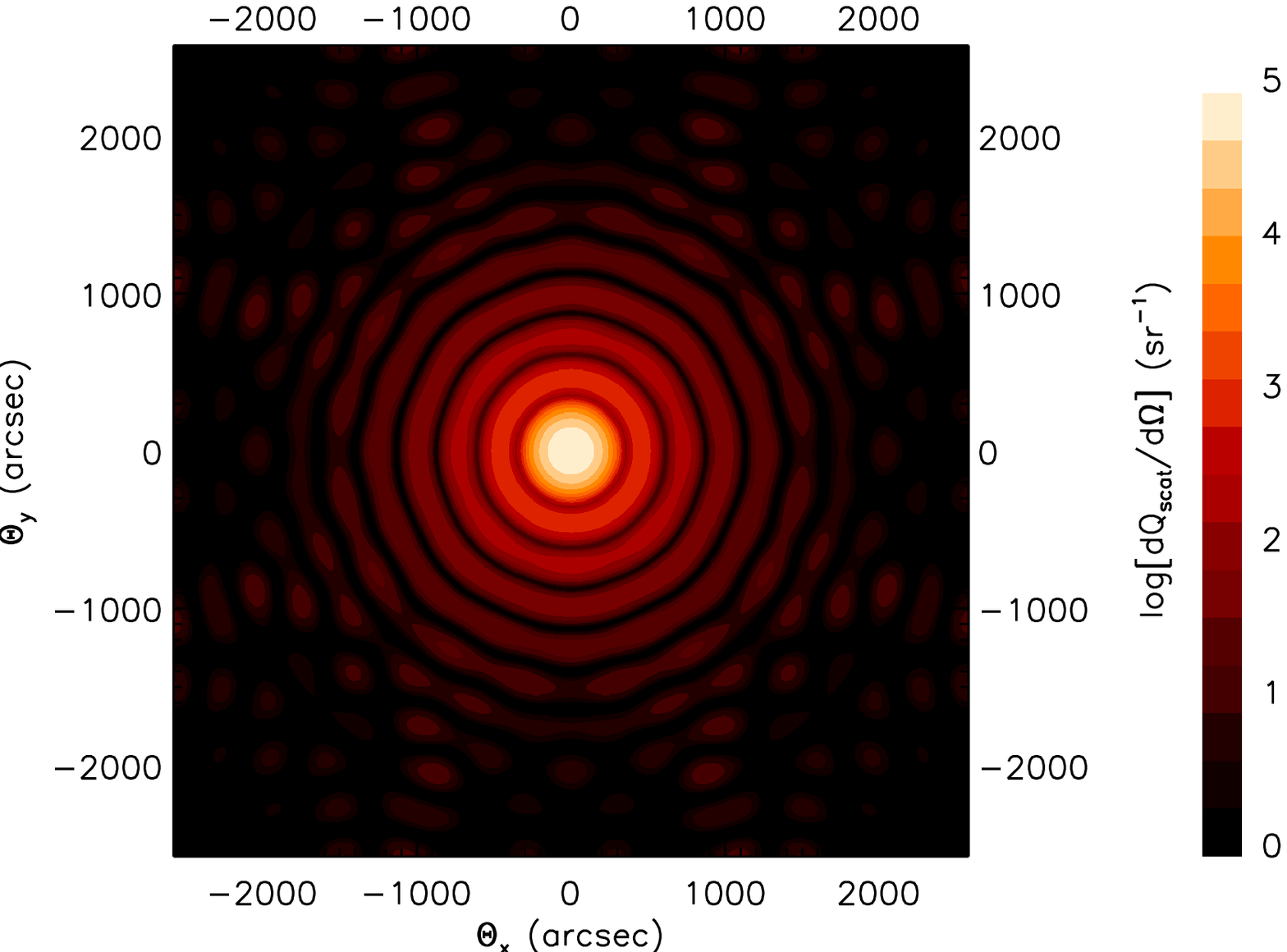}
\includegraphics[width=3.2in]{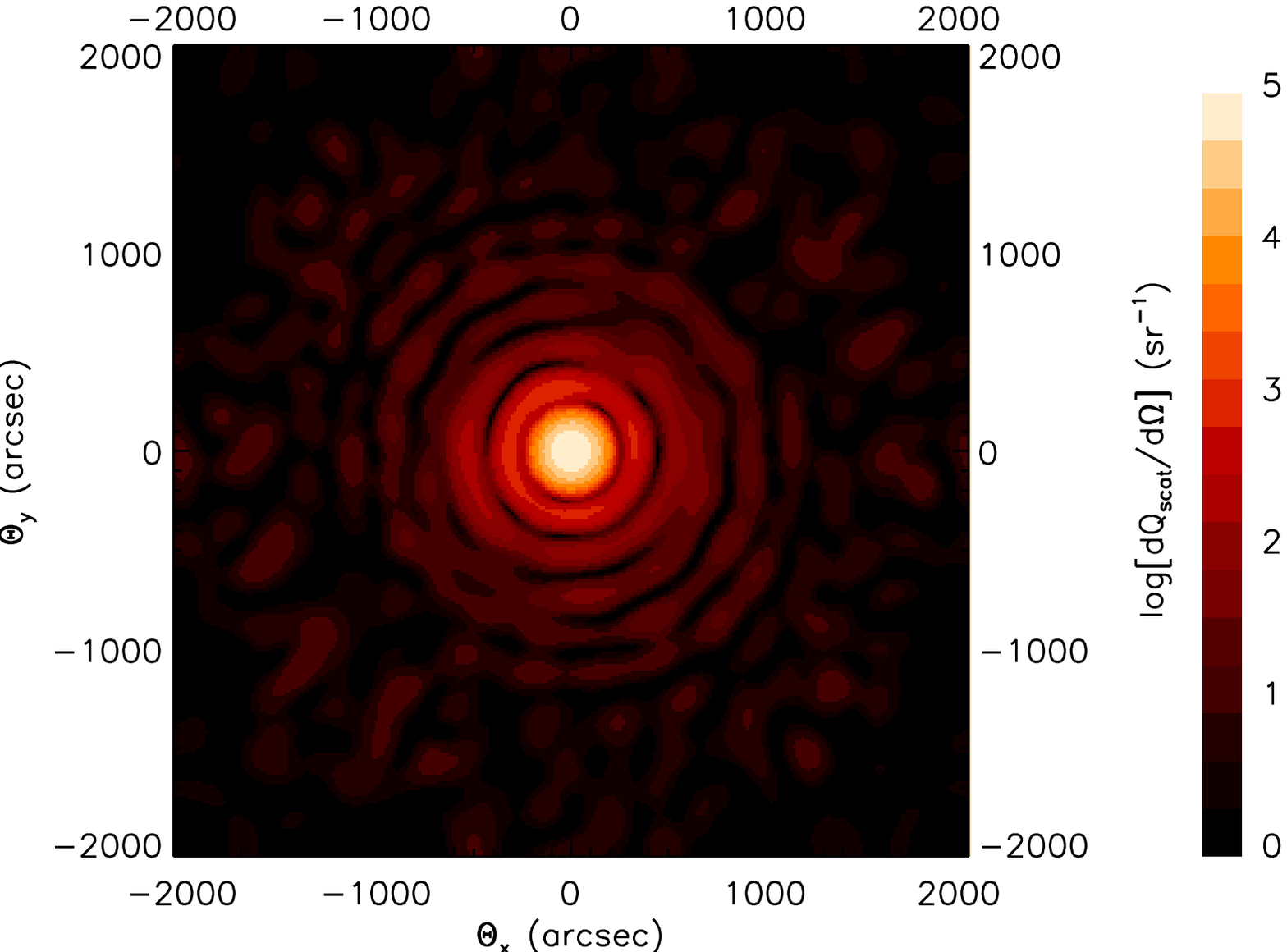}
\end{center}
\caption{Plots of $dQ_{\rm scat}/d\Omega$ ($E = 2$ keV) for (left) a
  single spherical grain with radius $a=0.25$ $\mu$m; and (right) the
  BAM2 aggregate of Fig.\ \ref{fig:schematic}, with $256$ $a_0 = 0.04$
  $\mu$m silicate spherules (${\cal P} = 0.563$, $a_{\rm eff} \approx
  0.254$ $\mu$m), oriented with $\theta = 0^\circ$, after
  ``$\beta$-averaging'' (see text).}
\label{fig:contour}
\end{figure}

\subsection{Anomalous Diffraction Theory (ADT)}

ADT, a combination of ray-tracing optics and Huygens' principle of
propagation (van de Hulst 1957), is applicable to the study of dust
grains at X-ray energies, when the refractive index is nearly unity
and the grain size is relatively large (van de Hulst 1957):
\begin{equation}
\begin{split}
&\vert m-1 \vert \ll 1,\\
&k_0 a \gg 1,\\
\end{split}
\end{equation}
where $m$ is the complex refractive index, $\lambda_0 = 2 \pi / k_0$
is the wavelength of the incident photon and $a$ is the size of the
dust grain.  
Draine \& Allaf-Akbari (2006, hereafter DA06) applied the ADT to
calculate X-ray scattering from non-spherical grains.
The interested reader is referred to DA06 or chapter 10
of van de Hulst (1957) for more details.

Let ${\bf k}_i=(0,0,k_0)$ and 
${\bf k}_s=(k_x,k_y,\sqrt{k_0^2-k_x^2-k_y^2})$ be the
incident and scattered propagation vectors.
The differential scattering cross section is given by
\begin{equation}
\frac{d\sigma_{\rm scat}}{d\Omega} = \frac{\vert {\cal S} \vert^2}{k^2_0},
\label{eq:diffscat}
\end{equation}
where the scattering function,
\begin{equation}
{\cal S}\left(k_x,k_y; k_0\right) = 
k^2_0 ~{\cal F}\left\{ f\left( x,y; k_0 \right) \right\},
\end{equation}
is related to the Fourier transform of the shadow function, $f=f(x,y; k_0)$:
\begin{equation}
{\cal F}\left\{ f\right\} \equiv 
\int ~\exp\left[i \left(k_x x + k_y y \right)\right] ~f ~dxdy.
\end{equation}

The shadow function quantifies the fractional change in the
propagating electric field, approximated as a plane wave, at a plane
located just beyond the grain:
\begin{equation}
f\left( x,y; k_0 \right) = 
1 - \exp{\left[ i k_0 \int \left(m-1\right) ~dz \right]},
\end{equation}
where $m=m(x,y,z)$ in general.

\subsection{Scaling Laws}
\label{subsect:scaling}

We define the ``effective
radius'' (or ``volume equivalent radius''),
\begin{equation}
a_{\rm eff} \equiv \left( \frac{3V}{4\pi} \right)^{1/3},
\label{eq:vradius}
\end{equation}
where $V$ is the volume of solid material in the grain.  If ${\cal P}$
is the grain porosity as defined by SDJ08, then the characteristic
size is
\begin{equation}
R \equiv a_{\rm eff} (1-{\cal P})^{-1/3}.
\end{equation}
The grain is assumed to be made up of $N_s$ spherical monomers
(``spherules'') with radii $a_0$ such that $a_{\rm eff}=N^{1/3}_s
a_0$.  The porosity ${\cal P}={\cal P}(N_s)$, depends on
the agglomeration scheme used (see \S\ref{sect:results}).  In
Fig.\ \ref{fig:schematic}, we show visualizations of three random aggregates.

Consider an X-ray photon with wave number $k_0$ encountering a dust
grain and deflected by an angle $\Theta$ and a transverse wave
number, $k_\perp = k_0 \sin{\Theta}$.  The characteristic scattering angle is
\begin{equation}
\begin{split}
\Theta_{\rm char} &\equiv \frac{2}{k_0 R} = \frac{\lambda_0}{\pi R} \\
&= 271^{\prime\prime} ~\left(1-{\cal P}\right)^{1/3} 
\left(\frac{\mbox{1 keV}}{E} \right)
\left(\frac{\mbox{0.3}\mu\mbox{m}}{a_{\rm eff}}  \right) ~~~.~~~\\
\end{split}
\label{eq:charangle}
\end{equation}
The total scattering cross section is
\begin{equation}
\sigma_{\rm scat} \approx \left(\frac{d\sigma_{\rm
    scat}}{d\Omega}\right)_{\Theta=0} ~\pi \Theta^2_{\rm char}.
\end{equation}

We can derive approximate scaling laws for $d\sigma_{\rm
  scat}/d\Omega$ at small and large angles.  At sufficiently high
energies such that $\vert f \vert \ll 1$, we have
\begin{equation}
{\cal F}\left\{ f \right\} \approx \sum_j \exp{\left(i \phi_j \right)} ~{\cal F}_0,
\end{equation}
where ${\cal F}_0$ is the Fourier transform of the shadow function for
one spherule and $\phi_j$ is the phase contribution by the $j$-th
spherule.

At small scattering angles ($k_\perp R \lesssim 1$; ``core''), we have
${\cal F}_0 \approx -ik_0 (m-1) V_0$ where $V_0 = 4\pi a^3_0/3$ is the
volume of one spherule.  The phase shifts are small
($\left<\phi^2_j\right> \lesssim 1$) and coherent scattering occurs:
\begin{equation}
{\cal F} \approx N_s {\cal F}_0.
\end{equation}
It follows that
\begin{equation}
\frac{d\sigma_{\rm scat}}{d\Omega} \approx \left( N_s V_0 k^2_0 \right)^2 \left\vert m-1 \right\vert^2.
\label{eq:12}
\end{equation}

At large scattering angles ($k_\perp R \gg 1$; ``wing''), the phase
shifts are large and pseudo-random so that
\begin{equation}
\left \vert {\cal F} \right \vert \approx 
N^{1/2}_s \left \vert {\cal F}_0 \right \vert.
\end{equation}
To evaluate ${\cal F}_0$, we examine the scattering function for a
single spherule (van de Hulst 1957):
\begin{equation}
{\cal S}_0\left(\Theta; k_0 \right) = \left(k_0 a_0\right)^2 \int^{\pi/2}_0 ~\left[ 1 - \exp{\left(-i\rho \sin{u}\right)} \right] {\cal J}_0\left(\chi \cos{u}\right) ~\sin{u} ~\cos{u} ~du,
\end{equation}
where $\rho \equiv 2k_0 a_0 (m-1)$, $\chi \equiv k_0 a_0 \Theta$ and ${\cal J}_0$ is the zeroth order Bessel function.  When $\chi \gg 1$ and $\rho \sim 1$, 
\begin{equation}
\begin{split}
{\cal S}_0\left(\Theta; k_0 \right) &\approx i \left(\frac{\pi}{2}\right)^{1/2} \left(k_0 a_0\right)^2 \rho \chi^{-3/2} {\cal J}_{3/2}\left(\chi\right)\\
&\approx 2i \left(m-1\right) k_0 a_0 \Theta^{-2},\\
\end{split}
\label{eq:sfunc_largek}
\end{equation}
since (Arfken \& Weber 1995)
\begin{equation}
{\cal J}_l\left(\chi \right) \approx \sqrt{\frac{2}{\pi \chi}} ~\cos{\left[\chi - \frac{\pi}{2}\left(l +\frac{1}{2}\right)\right]}
\end{equation}
for $\chi \gg 1$.  We then have
\begin{equation}
\frac{d\sigma_{\rm scat}}{d\Omega} \approx 4 N_s \left\vert m-1
\right\vert^2 a^2_0 \Theta^{-4} 
= 4|m-1|^2 \left(\frac{a_{\rm eff}^3}{a_0}\right) \Theta^{-4} ~~~.
\end{equation}
The scattering at large angles is proportional to the ``power''
at large $k_\perp$ in $|{\cal F}(f)|^2$.

Consequently, the normalized differential scattering cross section follows the scaling laws:
\begin{equation}
\frac{dQ_{\rm scat}}{d\Omega} \equiv \frac{1}{\pi a^2_{\rm eff}} \frac{d\sigma_{\rm scat}}{d\Omega} \propto
\begin{cases}
N^{4/3}_s &, \Theta \ll \Theta_{\rm char},\\
N^{1/3}_s \Theta^{-4} &, \Theta \gg \Theta_{\rm char}.\\
\end{cases}
\label{eq:scaling}
\end{equation}
We expect these scaling laws to be obeyed when absorption effects are weak:
\begin{equation}
\tau \equiv k_0 a_{\rm eff} m_i \left(1-{\cal P}\right)^{2/3} \ll 1,
\label{eq:tau}
\end{equation}
where $m_i$ is the imaginary part of the index of refraction.  For $E \ge 1$ keV, we have $m_i \le 1.9 \times 10^{-4}$ for olivine (MgFeSiO$_4$) and thus $\tau \le 0.3 (1-{\cal P})^{2/3}$ for $a_{\rm eff} = 0.3$ $\mu$m.

\subsection{Implementation}

To calculate the differential scattering cross section in equation
(\ref{eq:diffscat}), we adapt the {\tt Fortran 90} routine {\tt
  four2.f90} from Press et al. (1996) for computing two-dimensional
fast Fourier transforms (FFTs) with complex input functions.  We are
able to reproduce all of the results in Fig. 1 of DA06, thus verifying
the accuracy of our code.  As in the case of DA06, we find that a
square, cartesian grid $(x,y)$ with $(2^{11})^2$ to $(2^{12})^2$
points suffices for our purposes and avoids spurious contributions to
the FFT resulting from aliasing.  Bilinear interpolation is used to
map the calculations into polar coordinates.

\section{Results}
\label{sect:results}

%figure 3: ns=256 aggregate
\begin{figure}
\begin{center}
\includegraphics[width=3.5in]{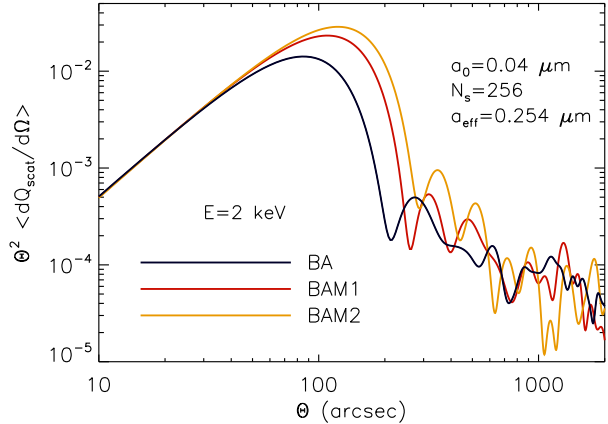}
\end{center}
\caption{$\left<dQ_{\rm scat}/d\Omega\right>$ at $E=2\,$keV as a
  function of the scattering angle, $\Theta$, for BA.256.1 (${\cal P}
  = 0.860$), BAM1.256.1 (${\cal P} = 0.706$) and BAM2.256.1 (${\cal P}
  = 0.563$) aggregates, each consisting of $256$ $a_0=0.04$ $\mu$m
  silicate spherules.}
\label{fig:256}
\end{figure}

%figure 4: scaling laws
\begin{figure}
\begin{center}
\includegraphics[width=3.5in]{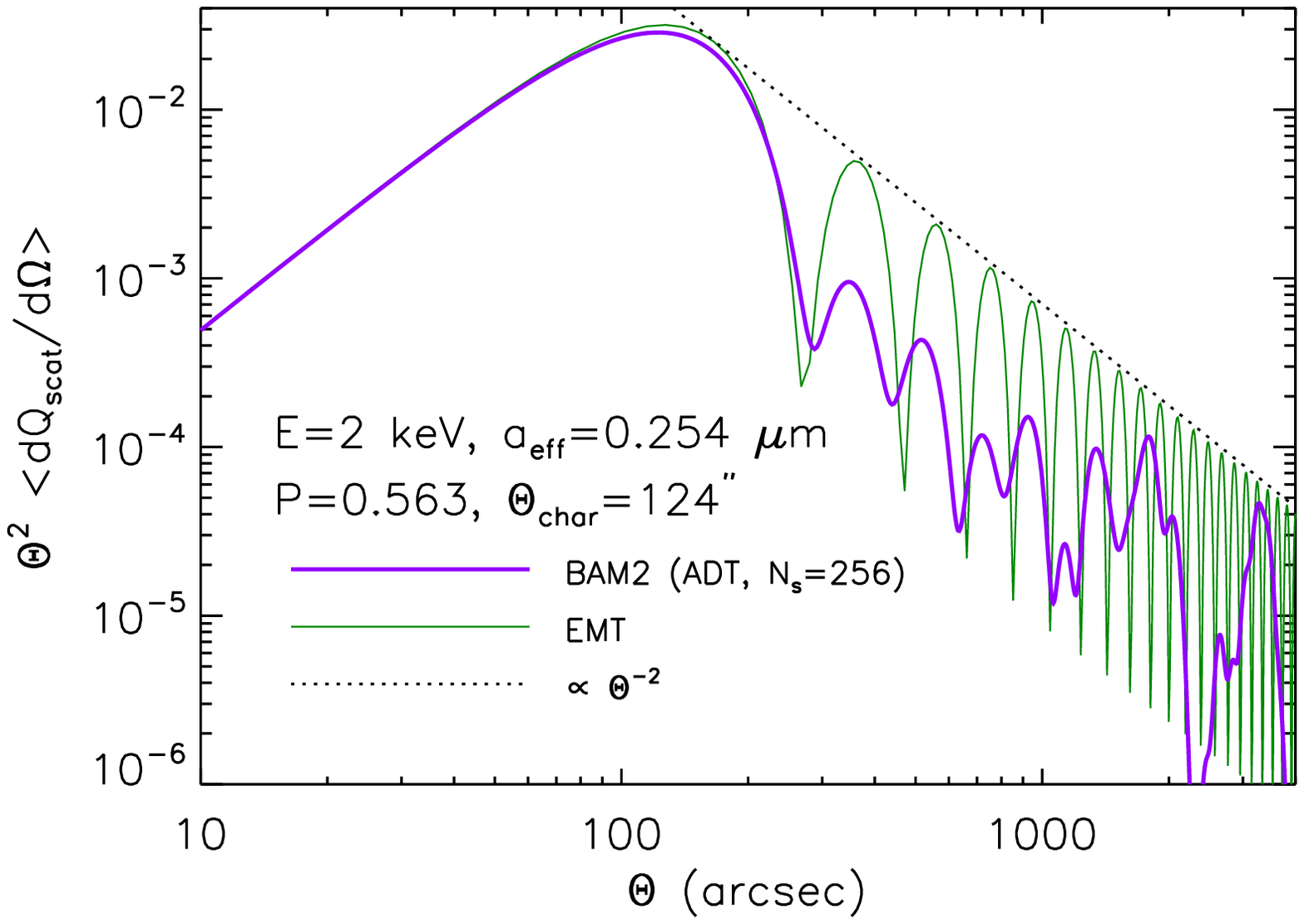}
\end{center}
\caption{Comparing ADT and EMT calculations for $\left<dQ_{\rm
    scat}/d\Omega\right>$ at $E=2\,$keV for BAM2.256.1 with
  $a_0=0.04\,\mu$m.}
\label{fig:adt_vs_emt}
\end{figure}
 
SDJ08 studied three types of grain aggregate models with different
ranges of porosities.  The most porous grains are the BA (``ballistic
agglomeration'') aggregates, constructed by requiring the arriving
monomers to adhere to the point where they first made contact with the
pre-existing aggregate.  Less porous are the BAM1 (``BA with one
migration'') aggregates, where an arriving monomer is required to roll
or slide along the contacted monomer, via the shortest possible
trajectory, until a second contact point is established with the
aggregate.  The most compact aggregates are the BAM2 (``BA with two
migrations'').  For $256 \le N_s \le 4096$, the BA, BAM1 and BAM2
aggregates have $\left<{\cal P}\right> \approx 0.85$--0.86, 0.74--0.78
and 0.58--0.66, respectively (see Table 2 of SDJ08).  The current
study employs random realizations of BA, BAM1, and BAM2
aggregates.\footnote{Available in electronic form at
  \texttt{http://www.astro.princeton.edu/$\sim$draine/agglom.html}.}  The
monomers are assumed to have composition MgFeSiO$_4$ and density
$\rho=3.8{\,\rm g\,cm}^{-3}$ of olivine, with refractive index from
Draine (2003).

The orientation of each aggregate is defined by two angles: $\theta$
is the angle between the line of sight and the principal axis of the
largest moment of inertia, with unit vector $\hat{a}_1$; and $\beta$
is the rotation angle about $\hat{a}_1$.  Rotation of $\hat{a}_1$
about the line of sight can be suppressed because the scattered
intensity will later be azimuthally averaged.  Because of the
reflection symmetry of the shadow function, we need to calculate
${\cal S}$ only for $0 \le \theta \le \pi/2$ and $0 \le \beta \le
\pi$.  For 11 values of $\theta$, we average $dQ_{\rm scat}/d\Omega$
over 11 values of $\beta$ (``$\beta$-averaging'').

We compute $dQ_{\rm scat}/d\Omega$ for the BAM2.256.1 aggregate
with $N_s = 256$ in the right panel of Fig.\ \ref{fig:contour}.  The
effective radius for this grain aggregate is $a_{\rm eff} =
0.254$ $\mu$m ($R \approx 0.335$ $\mu$m).  For comparison, we calculate
$dQ_{\rm scat}/d\Omega$ for a solid sphere with $a= 0.25$ $\mu$m.
The first feature to notice is the series of distinct rings in the case of
the solid sphere.  These are the two-dimensional analogue of the
Fourier transform of a Heaviside function in one dimension, and are
artifacts of the sharply-defined edge of the spherical grain.  For the
aggregate, traces of these rings persist, but they 
are smeared out because there is no sharp spherical outer edge.

Secondly, the scattering has similar intensities for both the
spherical grain and the grain aggregate, but the central peak is
concentrated within a smaller angular area for the latter.  This is a
simple consequence of the fact that porous grains generally have
larger characteristic sizes than solid spheres.  Larger grain sizes
result in more concentrated forward scattering.

For the rest of the paper, we average $dQ_{\rm scat}/d\Omega$ over 11
values of the orientation angle $\theta$ (``$\theta$-averaging'').  We
also define $\left<dQ_{\rm scat}/d\Omega\right>$ to be
the azimuthal average of the two-dimensional
$dQ_{\rm scat}/d\Omega$ function.  We show examples of $\left<dQ_{\rm
  scat}/d\Omega\right>$ for $N_s=256$ aggregates with BA, BAM1 and
BAM2 porosities in Fig.\ \ref{fig:256}.  As expected from equation
(\ref{eq:12}), the BA, BAM1, and BAM2 aggregates have nearly identical
forward scattering, but the BA clusters, being ``larger'', have a
narrower forward scattering lobe, and therefore weaker scattering at
intermediate angles (80--$200\arcsec$ in Fig.\ \ref{fig:256}).  At
large angles $\Theta\gtrsim500\arcsec$ the BA, BAM1, and BAM2 clusters
in Fig.\ \ref{fig:256} (all composed of the same number $N_s$ of
identical spherules) have similar $d\sigma_{\rm scat}/d\Omega$.  
%% We
%% checked that this trend is also observed for edge-on
%%  ($\theta=90^\circ$) vs. face-on ($\theta=0^\circ$) grains (without the
%% $\theta$-averaging procedure applied), due to the relative compactness
%% of the former.  
For grains with different values of $N_s$, we checked
the scaling laws described in equation (\ref{eq:scaling}) and verified
that they are obeyed to within a few percent.

In Fig. \ref{fig:adt_vs_emt}, we include as an example the effective
medium theory (EMT) calculation for a grain with the same mass and
porosity as a $N_s = 256$, BAM2 aggregate.  The EMT calculation
approximates the cluster by a uniform density sphere with refractive
index $m_{\rm EMT} = 1 + (1-{\cal P})(m-1)$.  For $\Theta \lesssim
\Theta_{\rm char}$, the EMT and ADT calculations are in agreement.  However, EMT does not take into account the
substructure of the grain aggregate, which is important for $\Theta \gtrsim \Theta_{\rm char}$.

We have also used the discrete dipole approximation (Draine \& Flatau 1994)
to calculate the extinction cross sections for the clusters
at the effective wavelengths of the $R$, $V$, and $B$ bands ($\lambda=0.6492$, 
$0.5470$,
and $0.4405 ~\micron$), using the public-domain
code DDSCAT 7.0 (Draine \& Flatau 2008).  Results for selected clusters
are shown in Table \ref{tab:targets}, together with the differential cross
section for forward-scattering of X-rays.

\section{X-Ray Scattering Diagnostics}
\label{sect:diagnostic}

%figure 5: diagnostic
\begin{figure}
\begin{center}
\includegraphics[width=3.5in]{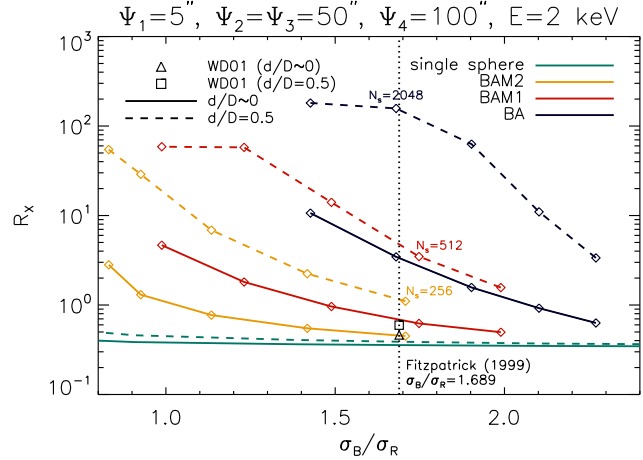}
\end{center}
\caption{$\RX$ vs. the ratio of $B$- to $R$-band extinction cross
  sections.  Higher $\RX$ and lower $\sigma_{B}/\sigma_{R}$ values
  correspond to larger grain sizes.  We have considered grain
  aggregates consisting of $N_s = 256$, 512, 1024, 2048 and 4096
  silicate spherules with radii $a_0=0.04$ $\mu$m.  For comparison,
  calculations for single silicate spheres are shown.  Also shown is
  the X-ray halo calculated for the WD01 mixture of silicate and
  carbonaceous grains.  The dotted vertical line is
  $\sigma_{B}/\sigma_{R}=1.689$ observed for typical diffuse cloud
  sightlines with $A_V/E(B-V)\approx3.1$ (Fitzpatrick 1999).  }
\label{fig:xray_vs_optical}
\end{figure}

% figure 6
\begin{figure}
\begin{center}
\includegraphics[width=3.5in]{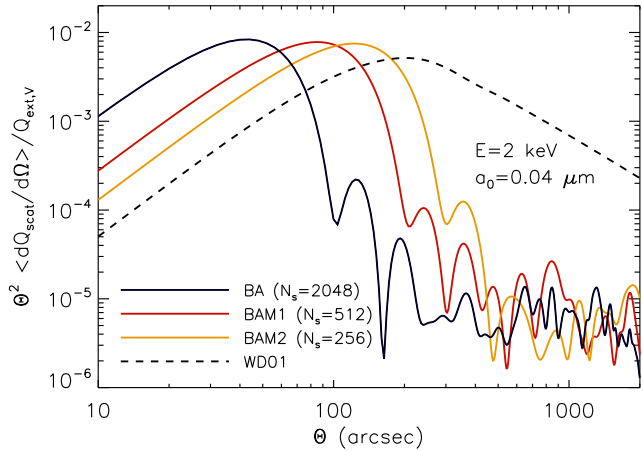}
\end{center}
\caption{\label{fig:compare} $[\Theta^2 dQ_{\rm sca}(2\,{\rm
      keV})/d\Omega]/Q_{\rm ext}(V\,{\rm band})$ for BA.2048.1--3,
  BAM1.512.1--3 and BAM2.256.1--3 aggregates (with $a_0=0.04$ $\mu$m)
  that approximately satisfy $\sigma_{B}/\sigma_{R}=1.689$ (see text).
  Forward scattering per unit $A_V$ increases with increasing porosity.
  The result for the WD01 dust model is also shown.
  }
\end{figure}

The scattering angle $\Theta$ is related to the observed halo angle
$\Psi$ via the relation:
\begin{equation}
d ~\tan{\Psi} = (D-d) ~\tan{\left( \Theta - \Psi \right)} ,
\end{equation}
where $d$ and $D$ are the distances to the dust population and X-ray
source, respectively.  For example, we have $\Psi = \Theta/2$ when
$d=D/2$.  Generally, we have $\Psi \approx \Theta$ when $d/D \ll 1$
and $\Psi \ll \Theta$ when $d \sim D$.

For $a_{\rm eff}\gtsim 0.1~\micron$ grains at optical wavelengths, the ratio
$\sigma_{B}/\sigma_{R}$ of extinction in the $B$- to the
$R$-band is a generally decreasing function of $a_{\rm eff}$
and serves as a diagnostic of the
grain size.  The observed value of $A_{B}/A_{R} \approx
1.689$ for the average interstellar reddening curve with
$A_V/E(B-V)\approx 3.1$ (Fitzpatrick 1999) requires
viable grain models to have the extinction at optical wavelengths
dominated by grains with $\sigma_{B}/\sigma_{R}\approx 1.689$.
Single-sized spheres with the dielectric function of
``astronomical silicate'' (Draine 2003) have
$\sigma_B/\sigma_R=1.689$ for radius $a=0.177\micron$.

A grain model can be tested by comparison with observed X-ray
scattering halos over a range of halo angles.  Because calculations of
optical-UV extinction and X-ray scattering by complex grain geometries
are very time-consuming, it is helpful to have simple diagnostics that
can be applied specifically to the grains that dominate the
extinction at optical wavelengths, because these same grains are expected
to dominate the total X-ray scattering at all except the largest angles.
Accordingly, we seek to characterize the X-ray scattering properties
of the grain model for scattering angles $\Theta \ltsim \Theta_{\rm char}(a=0.2\micron)\approx 407\arcsec\,({\rm keV}/E)$.

\subsection{Core Slope Diagnostic: $\cal{R}_{\rm X}$}
\label{subsect:rx}
For interstellar dust models based on compact grains, the optical
extinction is dominated by grains with radii $R\approx0.2\micron$, which
will have characteristic scattering angles
$\Theta_{\rm char}\approx 200\arcsec(2\,{\rm keV}/E)$.
For dust at $d/D\approx0.5$, the characteristic halo angle would be
$\Psi_{\rm char}\approx 0.5\Theta_{\rm char}$.
Here we devise a 
diagnostic $\RX$ that is based on the slope of the
scattering halo for $\Psi < \Psi_{\rm char}$:
\begin{equation}
\RX\left(\Psi_1,\Psi_2, \Psi_3, \Psi_4; E_1, E_2 \right)
\equiv \frac{\sigmax\left(\Psi_1,\Psi_2, E_1
  \right)}{\sigmax\left(\Psi_3,\Psi_4,E_2\right)},
\end{equation}
where
\begin{equation} \label{eq:sigma_X}
\sigmax\left(\Psi_a,\Psi_b,E\right) \equiv 
2\pi \int^{\Psi_b}_{\Psi_a} ~\frac{d\sigma_{\rm
    scat}}{d\Omega}\left(\Theta,E\right) ~\sin{\Psi} ~d\Psi.
\end{equation}
For fixed halo angles $\Psi_1 < \Psi_2 \leq \Psi_3 < \Psi_4$, we have
higher values of $\RX$ for larger grains because of the
increased forward scattering.  An integral property like ${\cal
  R}_{\rm X}$ provides a useful comparison between theory and
observations because it averages over oscillatory behavior in
$d\sigma_{\rm scat}/d\Omega(\Theta)$ when considering grains of a
single size.

For illustration, we set $E_1=E_2=2$ keV, and adopt the
following halo angles:
\begin{equation}
\begin{split}
&\Psi_1 = 5\arcsec ~~~\left(\ll \Theta_{\rm char}\right)\\
&\Psi_2 = 
\Psi_3 = 50^{\prime \prime} \left(\approx 0.25 \Theta_{\rm char} \right),\\
&\Psi_4 = 100\arcsec ~~~\left(\approx 0.5\Theta_{\rm char}\right)\\
\end{split}
\end{equation}
In Fig.\ \ref{fig:xray_vs_optical}, we construct curves of ${\cal
  R}_{\rm X}$ vs.\ $\sigma_{B}/\sigma_{R}$, for different
values of $d/D$.  The grain aggregates considered have $0.25 \lesssim
a_{\rm eff} \le 0.64$ $\mu$m and $0.33 \lesssim R \lesssim 1.25$
$\mu$m.  
%% As an example, the $N_s=2048$ and $N_s=4096$ aggregates have
%% $\sim 1$\% to 2\% differences in porosity, but may produce factors
%% $\sim 3$ differences in $\RX$.  
For the calculations in
Fig. \ref{fig:xray_vs_optical}, we have averaged the differential
cross sections over three random realizations of the grain
aggregates, a process we will incorporate for the rest of the paper.

The vertical dotted line in Fig.\ \ref{fig:xray_vs_optical} is the
``observed'' value of $\sigma_{B}/\sigma_{R}=1.689$ for dust
in the diffuse interstellar medium (ISM) with $R_V \equiv A_V/E(B-V) =
3.1$ (Fitzpatrick 1999).  It represents an average over the mixture of
grain sizes and composition present in the ISM.  For single-size
grains with a given porosity, $\sigma_{B}/\sigma_{R}$
decreases monotonically with increasing size for $\sigma_{\rm
  B}/\sigma_{\rm R} \gtrsim 1.2$.  The interstellar grain size
distribution extends over a wide range, but the grain size for which
$\sigma_{B}/\sigma_{R}=1.689$ will be characteristic of the
grains that dominate the extinction in the $B$-, $V$- and $R$-bands.

The computations reported for the aggregates were
CPU-intensive, therefore only a limited number of cases could be calculated.
In Fig.\ \ref{fig:compare}, we show the differential scattering cross
section for the BA ($N_s = 2048$), BAM1 ($N_s = 512$) and BAM2 ($N_s =
256$) grains that approximately satisfy the $\sigma_{\rm
  B}/\sigma_{\rm R}=1.689$ constraint.  We compare these calculations
to the scattering properties of the WD01 dust model.  
%% At scattering angles larger than about
%% $100\arcsec$, we see that our single-size approximation for the
%% spherules breaks down, as smaller grains are needed to produce X-ray
%% scattering at large angles comparable in intensity to the WD01
%% calculation.  This justifies the restriction that $\Psi_4 =
%% 100\arcsec$ (and not $\gg 100\arcsec$).

$\RX$ can be calculated for any grain model.  However, grain models
that are incompatible with the observed interstellar reddening law are
of no interest.  Restricting ourselves to aggregates that satisfy
$\sigma_{B}/\sigma_{R}=1.689$ determines $N_s$ for our chosen value of
$a_0 = 0.04$ $\mu$m.

%%We denote the value of $\RX$ that satisfies $\sigma_{\rm
%%  B}/\sigma_{\rm R}=1.689$ as
%% \begin{equation}
%% \tilde{{\cal R}}_{\rm X} \equiv 
%% \RX\left( \sigma_{B}/\sigma_{R}=1.689\right).
%% \end{equation}
%% For example, the BA, BAM1 and BAM2 aggregates shown in
%% Fig.\ \ref{fig:xray_vs_optical} have $\tilde{{\cal R}}_{\rm X}=3.38$,
%% 0.70 and 0.46, respectively, for $d/D \approx 0$.  

We compare our results in Fig.\ \ref{fig:xray_vs_optical} to
calculations for single astronomical silicate spheres with 
radii of $a=0.1$--0.3
$\mu$m.  The characteristic grain radius that satisfies $\sigma_{\rm
  B}/\sigma_{\rm R}=1.689$ is $a=0.177$ $\mu$m.  We also show
$\RX$ calculated for the WD01 size distributions
for carbonaceous (including polycyclic aromatic hydrocarbons) and
silicate grain populations for sightlines with $R_V=3.1$.  For $d/D
\le 0.9$, our single sphere calculations {\it underestimate} the
$\RX$ values calculated for the WD01 size
distributions, though by only $\sim$33\% on average.

Another issue to address is the grain composition adopted.  We have
neglected carbonaceous material, because the volume of silicate
material is $\sim$1.5 times the volume of carbonaceous solids.
Moreover, the silicates are more effective for X-ray scattering at $E
> 1$ keV because $\vert m-1 \vert$ is larger than for graphite (see
Figs. 3 and 5 of Draine [2003]).  At $E=2$ keV, $\vert m-1 \vert =
1.94 \times 10^{-4}$ for MgFeSiO$_4$ versus $1.13 \times 10^{-4}$ for
graphite; the scattering power scales as $\vert m-1 \vert^2$.
Therefore, the graphite grains make only a secondary contribution to
the scattering at $E>1$ keV.

\subsection{Core X-Ray Scattering per $A_V$ Diagnostic: $T_X$}
\label{subsect:tx}

We construct a second diagnostic based on the strength of the X-ray scattering
in the ``core'' relative to the optical extinction.
Define
\begin{equation} \label{eq:define TX}
\begin{split}
\TX &\equiv \left(\frac{D}{D-d}\right)^2 \frac{\Delta\tau_{\rm scat}
  \left(\Psi_1,\Psi_2,E \right)}{A_V}\\ &= 0.92
\left(\frac{D}{D-d}\right)^2
\frac{\sigmax\left(\Psi_1,\Psi_2,E\right)}{\sigma_V},\\
\end{split}
\end{equation}
where $\Delta\tau_{\rm scat} = (n_d/n_{\rm H}) N_{\rm H}\sigmax$,
$(n_d/n_{\rm H})$ is the number of dust grains per H nucleon, $N_{\rm
  H}$ is the column density, and $\sigmax$ is defined in equation
(\ref{eq:sigma_X}).  $\TX$ can be calculated for any mixture of
grains, but we are again only interested in grain models that are
compatible with the observed interstellar reddening law.  The
forward-scattering cross section is strongly dependent on the grain
size, and therefore this diagnostic is sensitive to the presence of
larger grains in the size distribution.

\subsection{Comparison to Measured Dust Haloes}
\label{sect:comparison}

%figure 7: GX+13 R_X
\begin{figure}
\begin{center}
\includegraphics[width=3.5in]{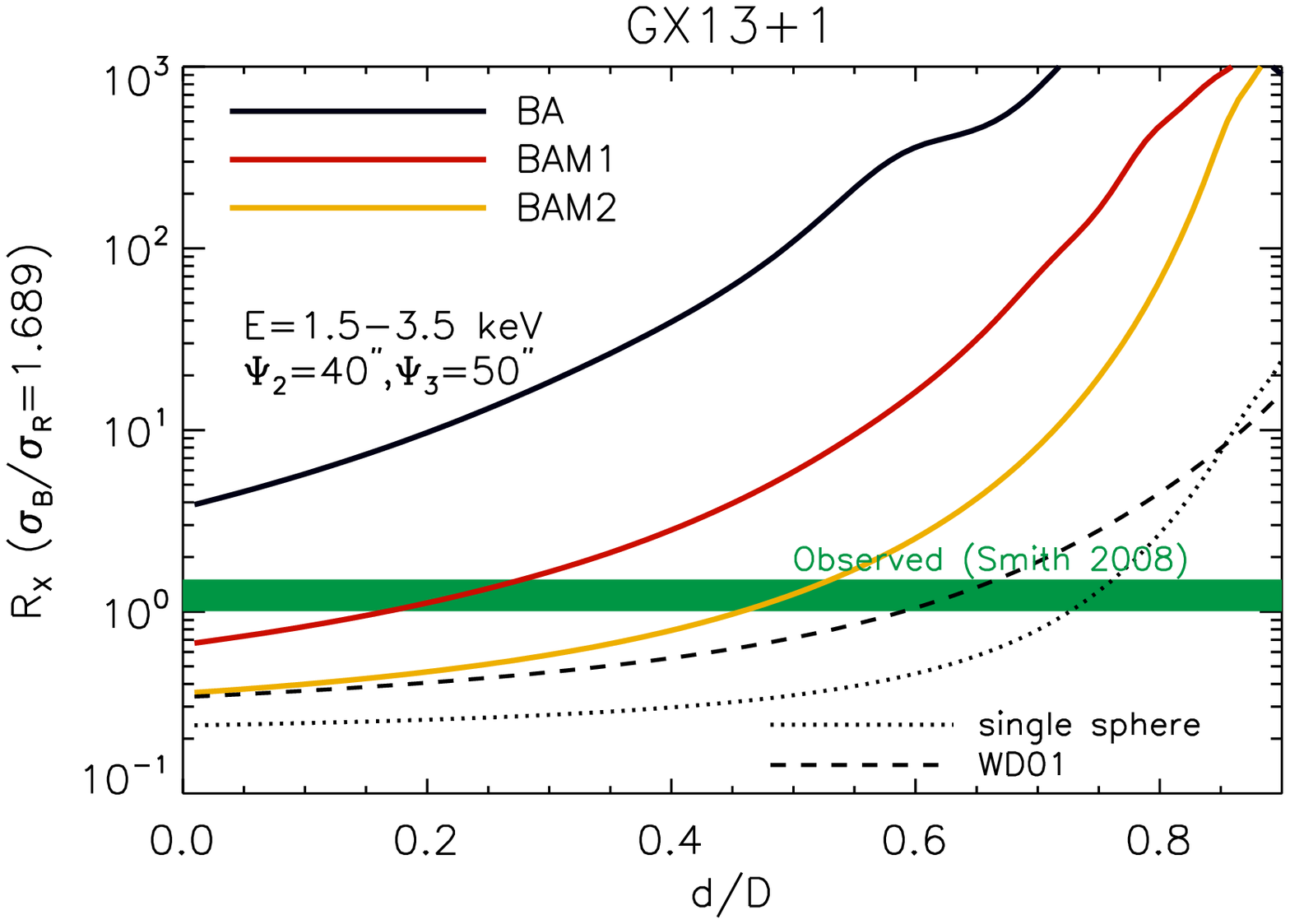}
\end{center}
\vspace*{-0.2cm}
\caption{\label{fig:gx13_rx} Observed and theoretical $\RX$ vs.\ the
  ratio of the distances to the dust population and GX13+1.  The width
  of the ``observed'' band reflects uncertainties in using both the
  HRC-I and ACIS-I data to evaluate
  $\sigmax\left(50^{\prime\prime},100^{\prime\prime},E\right)$.}
\vspace*{0.2cm}
\end{figure}

%figure 8: GX+13 T_X
\begin{figure}
\begin{center}
\includegraphics[width=3.5in]{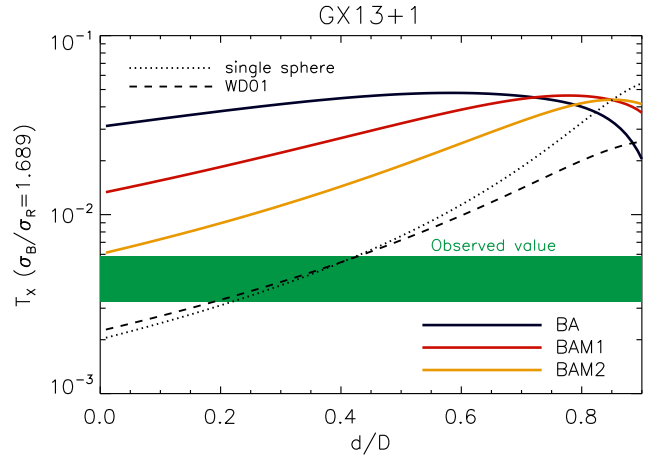}
\end{center}
\caption{\label{fig:gx13_tx}
         Observed and theoretical $\TX$ vs.\ the
  ratio of the distances to the dust population and GX13+1.
  The width of the ``observed'' band reflects uncertainties in $A_V$.
  }
\end{figure}

The $\RX$ and $\TX$ 
diagnostics constructed in
\S\S\ref{subsect:rx},\ref{subsect:tx} allow us to compare our theory with
observations of X-ray dust haloes.  Recently, high-precision {\it
  Chandra} observations of scattered haloes around X-ray binaries have
allowed for detailed comparisons to grain models (S08; Thompson \&
Rothschild 2009).  S08 gives a flavor of the technical difficulty of
the X-ray background subtraction involved in measuring the dust halo
around the Galactic binary GX13+1 using HRC-I and ACIS-I.  The HRC-I
data is measured for $2^{\prime\prime} \lesssim \Psi \lesssim
786^{\prime\prime}$, while the ACIS-I data is for $50^{\prime\prime}
\lesssim \Psi \lesssim 989^{\prime\prime}$.  S08 noted the good level
of agreement between the HRC-I and ACIS-I data between $50^{\prime
  \prime}$ and $100^{\prime \prime}$.  Uncertainties with background
subtraction make it difficult to study X-ray halos at large halo
angles ($\Psi\gtrsim500\arcsec$).  At small angles
($\Psi\lesssim5\arcsec$), it becomes difficult to untangle the halo
signal from the instrumental point spread function (resulting from,
e.g., micro-roughness of the telescope mirror).

The range of halo angles available in the GX13+1 measurement makes it
an ideal case study for us.  However, the HRC-I count rate is
dominated by the point-spread function (PSF) for $\Psi \gtrsim
200^{\prime \prime}$ (see \S3 of S08) and the measured PSF provided to
us by R.K. Smith (2008, private communication) in electronic form only
extends up to about $180^{\prime\prime}$.  Since the HRC-I has no
energy resolution, S08 used the best-fit model for the measured {\it
  RXTE} PCA spectrum of GX13+1 and folded it through the HRC-I
response function (see Fig. 3 of S08).  We compute the effective
$\sigmax\left(\Psi_1,\Psi_2, E \right)$ by taking an average over the
energy spectrum of detected counts:
\begin{equation}
\overline{\sigmax}
\left(\Psi_1,\Psi_2, E \right) = \sum_{j=1}^4 W_{E_j}
\sigmax\left(\Psi_1,\Psi_2, E_j \right),
\end{equation}
where the index $j$ runs over $E_j=1.75$---3.25 keV in intervals of
$\Delta E = 0.5$ keV.  The characteristic energy is then $\bar{E}
\approx 2.5$ keV; we set $\Psi_1 = 4\arcsec$ and $\Psi_2 = 40^{\prime
  \prime}$.  The weights are
\begin{equation}
W_{E_j} = \frac{\int^{E_j + \Delta E/2}_{E_j - \Delta E/2}
  ~N_E ~dE}{\int_ E ~N_E ~dE},
\end{equation}
where $N_E$ is the energy spectrum of detected counts.  Similarly, we
compute $\overline{\sigmax}\left(\Psi_3,\Psi_4, E \right)$ by taking
$\Psi_3 = 50\arcsec$ and $\Psi_4 = 100^{\prime \prime}$.  Using
$\overline{\sigmax}\left(\Psi_1,\Psi_2, E \right)$ and
$\overline{\sigmax}\left(\Psi_3,\Psi_4, E \right)$, we construct
theoretical values of $\RX$ in Fig. \ref{fig:gx13_rx}.

% Since the ACIS-I data is only measured for $\Psi \gtrsim 50\arcsec$, we take
%$\Psi_3 = 50^{\prime \prime}$.  Following the discussion in
%\S\ref{subsect:rx}, we take $\Psi_4 = 100\arcsec$.  We compute
%$\sigmax\left(\Psi_3,\Psi_4, E_2 \right)$ using $E_2 = 2.5$ keV, since
%the ACIS-I data is for $2.5 \pm 0.1$ keV.  Using
%$\sigmax\left(\Psi_1,\Psi_2, E_1 \right)$,
%$\sigmax\left(\Psi_3,\Psi_4, E_2 \right)$ and $\sigma_{\rm ext}(V)$,
%we construct theoretical values of $\RX$ and $\TX$, which we show in
%Figs.\ \ref{fig:gx13_rx} and \ref{fig:gx13_tx}.
 
Using the HRC-I and ACIS-I data and the stated halo angles yields the
measured value of $\RXobs=1.26 \pm 0.25$.  The theoretical value
depends on the assumed value of $d/D$.  From Fig.\ \ref{fig:gx13_rx},
we see that the BA model appears to be ruled out by the $\RX$
test.  The BAM1 and BAM2 models are compatible with the $\RX$ test for $d/D \approx 0.25$ and 0.5, respectively.

We show theoretical values of $\TX$ in Fig. \ref{fig:gx13_tx}.  To
obtain $\TXobs$, we need $A_V$.  Ueda et al. (2004) measured the
column density along the line of sight to GX13+1 to be $N_{\rm H} =
(3.2 \pm 0.2) \times 10^{22}$ cm$^{-2}$.  Since GX13+1 is located
towards the inner galaxy, we assume $A_V/N_{\rm H}$ to be
$1.25\pm0.25$ times the local value ($4.65 \times 10^{-22}\,{\rm
  mag\,cm}^2$; Rachford et al.\ 2009) and obtain $A_V = 18.6 \pm 3.9$,
somewhat larger than earlier estimates $A_V=10.2\pm 0.3$ (Garcia et
al.\ 1992) and $A_V=15.4\pm2.2$ (Charles \& Naylor 1992). The observed
value $\TXobs = 0.0046 \pm 0.0013$ is inconsistent with the BA and
BAM1 models for all values of $d/D$, but may be marginally compatible
with the BAM2 model provided $d/D\ltsim 0.1$.  However, it is clear
that the large extinction toward GX13+1 is not located at $d/D\approx
0$, but instead almost certainly arises mainly in the 3.5~kpc ring at
$d\approx 5\,$kpc where the Norma and Crux-Scutum arms cross the
sightline to GX13+1.  Therefore we conclude that all 3 fluffy-grain
models -- BA ($\poro\approx0.87$), BAM1 ($\poro\approx 0.73$), and
BAM2 ($\poro\approx 0.57$) -- are ruled out by $\TXobs$ toward GX13+1.
Based on both diagnostics, the porosity of interstellar dust is
$\poro\lesssim\poromax$.

%\btdnote{Kevin: With the most up-to-date determination of $A_V/N_H$,
%  lowering somewhat the estimated increase of $A_V/N_H$ toward the
%  inner galaxy, and allowing for uncertainty in this correction,
%  $\TXobs$ is increased from your earlier estimate by a factor
%  (23.9/18.6)=1.3, and the uncertainty range is also increased.} 

What do the $\RX$ and $\TX$ diagnostics say about compact grain
models?  The WD01 model, based on a mixture of compact silicate and
carbonaceous grains, reproduces $\RXobs$ if $d/D\approx 0.60-0.65$,
but reproduces $\TXobs$ only if $d/D \approx 0.2$--0.4.  In reality,
the dust is presumably distributed along the sightline to GX13+1
rather than residing in a single cloud.  From the variation of $\TX$
with $d/D$ in Fig. \ref{fig:gx13_tx} it is clear that WD01 dust
uniformly-distributed along the sightline would result in a $\TX$
value considerably larger than $\TXobs$, consistent with the
conclusion of Smith (2008) who found that smoothly-distributed 
WD01 model dust
would approximately reproduce the observed X-ray
scattering only if $N_{\rm H}\approx (1.5-2)\times10^{22}{\,\rm
  cm}^{-3}$, well below 
$N_{\rm H}\approx 3.2\times10^{22}{\,\rm cm}^{-2}$ estimated by Ueda
et al.\ (2004).  This may indicate that the WD01 dust model produces
too much X-ray scattering.  However,
% the WD01 model is
% consistent with X-ray scattering observed toward Nova Cygni 1992
%(Draine \& Tan 2003).  
(1) there may be systematic errors in the modeling
introduced by the broad spectrum of X-ray energies contributing to the
HRC-I imaging, and (2) the effects of multiple scattering are not 
negligible at the lowest energies.  In addition, the estimate of $A_V$
is based on X-ray absorption along the 
line of sight to GX13+1, whereas the X-ray
scattering at $300\arcsec$ is produced by dust that may be displaced
from the line of sight by up to $\sim$10 pc.  Perhaps $A_V$ on 
the direct sightline to
GX13+1 is larger than the average dust column
contributing to the scattered halo.  For example, if the dust producing the
scattering halo toward GX13+1 had $A_V\approx 12\pm 3$ rather
than $18.6\pm3.9$, the ``observed'' band in Fig.\ \ref{fig:gx13_tx}
would shift upward by about a factor of 1.55, and the WD01 model would
be more-or-less consistent with both $\RXobs$ and $\TXobs$ for
$d/D\approx\poromax$.  Compact grain models therefore appear to be
viable, whereas all three of the fluffy grain models considered here
appear to be firmly ruled out by the $\TX$ diagnostic.\footnote{If
  $A_V\approx 10$, the BAM2 model would be allowed by the $\TX$
  diagnostic if $d/D\lesssim 0.24$, but the $\RX$ diagnostic requires
  $d/D\approx 0.50$, so the BAM2 model would remain incompatible with
  the observational constraints.}

%We note that S08 was able to fit the WD01 model to HRC-I and ACIS-I data: 1. if $d/D < 10^{-3}$; 2. by assuming that
%the dust had both a smoothly-distributed and ``cloud'' component.  They also note that
%models using only a single thin cloud (with no smooth component) gave generally poor fits independent of the dust model used,
% and did not quote the derived values of $d/D$ from these fits.

% These results
%are consistent with the model fits of S08, where $d/D \sim 0$ was
%required in order for the ``single cloud'' component to work.
%\btdnote{Kevin: I'm not sure the statement claiming agreement with the
%  conclusions of S08 regarding WD01 will survive the increase in
%  $\TXobs$.}  and \ref{fig:gx13_tx} we see that there is no value of
%$d/D$ for which the theoretical values predicted by the BA, BAM1, and
%BAM2 models are consistent with the measured one, effectively setting
%an upper limit on the grain porosity ${\cal P}\ltsim\poromax$.

\section{Discussion}
\label{sect:discussion}

We have shown how the global structure of porous dust grain agglomerates
leads to increased X-ray scattering at small angles, when compared
with the scattering properties of spherical grains (both single size
and the WD01 size distribution) that are constrained to have the
observed ratio of extinction in the $B$ and $R$ bands.  The enhanced
small-angle scattering implies that observations of scattered X-ray halos 
can be used to
constrain the internal geometry of interstellar grains.
Applying the $\RX$ and $\TX$ diagnostics described above, we find that
the observed X-ray halo around GX13+1 is inconsistent with
grains consisting of random aggregates with porosities ${\cal P}
\gtrsim\poromax$, at least for the monomer sizes considered here
($a_0=0.04\,\mu$m).  

Smith et al.\ (2002) argued that the dust toward GX13+1 could not
consist of porous grains, on the grounds that porous grains could not
produce sufficient overall scattering.  Our $\TX$ diagnostic uses the
strength of the ``core'' of the scattered halo to constrain the dust
porosity, but now we use the fact that porous grains would produce
{\it too much} small-angle scattering per unit $A_V$.

This exploratory study has focused on the X-ray scattering properties of only 
the grains that dominate the optical extinction, with the expectation that
these are representative.
Future studies should: 
\begin{enumerate}
\item Use size distributions of grain aggregates (adjusted to
  reproduce the full extinction curve and not just $A_B/A_R$) rather
  than just averaging over random realizations of a single aggregate
  size as was done here.  We have shown that the $\RX$ and $\TX$
  diagnostics for single-size silicate spheres with
  $\sigma_B/\sigma_R=1.69$ provide a good approximation to $\RX$ and
  $\TX$ calculated for the full WD01 size distribution, so we are
  confident that using single-size fluffy grains with
  $\sigma_B/\sigma_R\approx1.69$ provides a reasonable approximation
  to what $\RX$ and $\TX$ would be for size distributions constrained to
  reproduce the observed extinction curve.
  Nevertheless,  
  $\RX$, in particular, will
  change when the single ``representative'' grain size is replaced by
  a size distribution that reproduces the IR-UV extinction curve.
  The $\TX$ diagnostic is expected to be more robust, because
  the grains that dominate the visual extinction are also expected to
  dominate the scattering at $\Theta < \Theta_{\rm
    char}(R=0.2\micron)$.
\item Include carbonaceous material, even though that material
  contributes only $\sim$40\% of the total grain volume.
  Again, we have already tested for sensitivity to the carbonaceous
  component by showing that $\RX$ and $\TX$ for single-size silicate spheres
  approximates $\RX$ and $\TX$ for the full WD01 size distribution, 
  including carbonaceous grains, but it would be best to explicitly
  consider fluffy grain models that include a carbonaceous component.
\item Examine the effect of varying the monomer size.  The monomer
  size used in the present work (0.040$\,\mu$m) was chosen
  arbitrarily.  While we expect that similar results will be obtained
  for monomer sizes $a_0 \gtrsim 0.010\,\mu{\rm m}$, this remains to
  be verified by future (numerically challenging) calculations.
\end{enumerate}

%\btdnote{Kevin: As I already mentioned in email sent on 5/23, there is
%  one issue that continues to trouble me.  Smith 2008 states that the
%  WD01 model works if the column density $N_{\rm H} =
%  (1.5-2)\times10^{22}{\,\rm cm}^{-2}$.  Whereas we are claiming,
%  based on the $\TX$ test, that the WD01 model works OK for $N_{\rm
%    H}=3.2\times 10^{22}{\,\rm cm}^{-2}$ with an enhanced dust/gas
%  ratio.  How can these statements -- which appear to differ at the
%  factor-of-two level -- both be correct?  SES02 and S08 both used the
%  Rayleigh-Gans approximation, which will overestimate the scattering
%  when the absorption optical depth across the grain is not small, and
%  I wonder whether this may have led them to overpredict the X-ray
%  scattering per unit H for the WD01 model.  Or are we making a
%  mistake somewhere?}

Although confirmation by more extensive future studies (as discussed above)
will be invaluable, 
the diagnostics developed here, in particular the $\TX$ diagnostic,
already demonstrate that the dust toward GX13+1 has porosity
$\poro\ltsim\poromax$.  
This finding appears to rule out models (e.g., Mathis \& Whiffen
1989; Voshchinnikov et al.\ 2006) in which the interstellar grain mixture
is dominated by highly-porous aggregates with ${\cal P}\gtrsim 0.8$.

The porosity of interstellar dust grains therefore appears to be {\it
  lower} than the porosity of the submicron grains in the debris disk
around AU Mic, which (based on interpretation of observed polarized
scattering) have $\poro\approx0.6$, or micron-sized cometary dust
grains, which also appear to have $\poro\approx 0.6$ (Shen et
al.\ 2009, and references therein).

Why interstellar grains are {\it not} highly-porous is an open
question.  Gyroresonant acceleration by MHD turbulence in diffuse
clouds appears able to drive $a\gtrsim0.1\micron$ grains to velocities
$\sim1{\,\rm km\,s}^{-1}$ (Yan et al.\ 2004).  Grain-grain collisions
at relative velocities $\sim{\rm km\,s}^{-1}$ will likely shatter
high-porosity aggregates, and thereby limit the abundance of fragile
high-porosity aggregates in the ISM.

%%%%%%%%%%%%%%%%%%%%%%%%%%%%%%%%%%%%%%%%%%%%%%%%%%%%%%%%%%%%%%%%%%%%%%%%%
%\input table.tex
\begin{table}
\begin{center}
\footnotesize
\caption{\label{tab:targets}
         Optical Extinction and X-Ray Forward Scattering for Selected Targets}
\begin{tabular}{l c c c c c c c}
target & ${\cal P}$ & $a_{\rm eff}$ & $R$ & $Q_{\rm ext}$ &
                                            $Q_{\rm ext}$ &
                                            $Q_{\rm ext}$ &
                                            $(dQ_{\rm sca}/d\Omega)_0$ \\
       &            & $(\micron)$   & $(\micron)$  & R & V & B & 2~keV\\
\hline
BA.2048.1   & 0.8644 & 0.508 & 0.989 & 4.986 & 6.484 & 8.352 & $3.353 \times 10^6$\\
BA.2048.2   & 0.8657 & 0.508 & 0.992 & 4.958 & 6.460 & 8.370 & $3.384 \times 10^6$\\
BA.2048.3   & 0.8651 & 0.508 & 0.990 & 5.004 & 6.502 & 8.392 & $3.364 \times 10^6$\\

BAM1.512.1  & 0.7239 & 0.320 & 0.491 & 3.358 & 4.425 & 5.764 & $5.319 \times 10^5$\\
BAM1.512.2  & 0.7530 & 0.320 & 0.510 & 3.211 & 4.308 & 5.749 & $5.326 \times 10^5$\\
BAM1.512.3  & 0.7286 & 0.320 & 0.494 & 3.331 & 4.398 & 5.789 & $5.279 \times 10^5$\\

BAM2.256.1  & 0.5632 & 0.254 & 0.335 & 2.877 & 3.771 & 4.877 & $2.116 \times 10^5$\\
BAM2.256.2  & 0.5781 & 0.254 & 0.339 & 2.841 & 3.736 & 4.822 & $2.103 \times 10^5$\\
BAM2.256.3  & 0.5818 & 0.254 & 0.340 & 2.807 & 3.732 & 4.857 & $2.085 \times 10^5$\\

sphere      & 0      & 0.177 & 0.177 & 2.499 & 2.940 & 4.227 & $5.012 \times 10^4$\\
\hline
\multicolumn{8}{l}{All targets consist of astronomical silicate (Draine 2003).}
\end{tabular}

\end{center}
\end{table}
%-------------------------------------------------------------------------

\acknowledgments \scriptsize K.H. acknowledges generous support by the
Institute for Advanced Study, especially for the use of the computing
cluster that served as the untiring workhorse for this project.  B.T.D. acknowledges partial support
by NSF grant AST 0406883.  We thank Mario Juric for helpful technical advice, Bernd Aschenbach
for an illuminating conversation about X-ray mirrors, and Randall Smith for providing us with his dust halo measurements in
an electronic form.  \normalsize

%%%%%%%%%%%%%%%%%%%%%%%%%%%

%REFERENCES

\end{document}